\documentstyle[sprocl]{article}

\input{epsf}

\begin{document}

{\noindent\small UNITU--THEP--5/1999 \hfill FAU--TP3--99/3 
\hfill hep-ph/9905324 }\vspace{5mm}

\title{The Infrared Behavior of Propagators in Landau Gauge QCD\footnote{
Talk given by Reinhard Alkofer at the
Workshop "Understanding Deconfinement in QCD", ECT*, Trento, 1-13/March, 1999.}}

\author{Reinhard Alkofer, Steven Ahlig}

\address{Universit\"at T\"ubingen, Institut f\"ur Theoretische Physik,\\
         Auf der Morgenstelle 14, 72076 T\"ubingen, Germany\\
	 E-mail: Reinhard.Alkofer@uni-tuebingen.de, \hbox{\hskip 1cm}\\
           \hskip 1cm ahlig@pion06.tphys.physik.uni-tuebingen.de}

\author{Lorenz von Smekal}
\address{Universit\"at Erlangen--N\"urnberg, 
	 Institut f\"ur Theoretische Physik III,\\
	 Staudtstr.~7, 91058 Erlangen, Germany\\
	 E-mail: smekal@theorie3.physik.uni-erlangen.de}

\maketitle	 
	 

\begin{abstract}
A closed system of equations for the propagators of Landau gauge QCD is
obtained in a truncation scheme for their Dyson--Schwinger equations
which implements the Slavnov--Taylor identities for the 3--point
vertex functions while neglecting contributions from irreducible 
4--point correlations. In the pure gauge theory without quarks, 
non--perturbative solutions for the gluon and ghost propagators 
are available in an approximation which allows for an analytic
discussion of their behavior in the infrared: The gluon propagator vanishes
for small spacelike momenta, whereas the ghost propagator is found
to be infrared enhanced. The running coupling of the non--perturbative
subtraction scheme approaches the finite value $\alpha_c \simeq 9.5$
at an infrared fixed point. The gluon propagator entails a violation of
positivity for transverse gluon states implying their absence 
from a positive subspace expected for asymptotic hadronic states
and thus confinement.
Both propagators, obtained for gluons and ghosts in the present scheme, 
compare well with recent lattice calculations. 

In the quenched approximation, the quark propagator describes dynamical
chiral symmetry breaking well, although the corresponding interaction in
the gap equation for the quark self--energy is infrared suppressed. 
First results of a simultaneous solution to the 
coupled system of gluons, ghosts {\sl and} quarks indicate towards weak, and
possibly negligible, vacuum polarisation effects of dynamical quarks on
the gluon and ghost correlations in the infrared.
\end{abstract}

\section{Introduction}

Regarding its phenomenological implications on hadron physics,\cite{Tan97}
and on ultrarelativistic heavy ion collisions, the infrared behavior of QCD
Green's functions comprises the essential information to describe  
non--perturbative phenomena. This infrared behavior should not only reflect
confinement and chiral symmetry breaking, but also the respective
deconfinement and chiral transition(s) at finite temperature and
density. In Landau gauge QCD at zero temperature and density,
our knowledge of the elementary correlations, the gluon, ghost and quark
propagators, in the infrared has recently made progress. Corresponding
solutions of truncated Dyson--Schwinger equations
(DSEs)\cite{Sme97,Sme98,Atk97,Atk98} and presently 
available results from lattice simulations\cite{Sum96,Der98,Attilio} 
now seem to yield a quite coherent picture.               
Models for these propagators at finite temperature and
density to describe the deconfinement and chiral transitions are currently
being developed.\cite{Trento}  

An important issue, remaining to be adressed, is the interrelation of
the different models of confinement that arise in different gauges. 
Unavoidable gauge ambiguities in gauges such as the Polyakov
or the maximal Abelian gauge seem to suggest pictures ranging from monopole
condensation and the dual Meissner effect to the dominance of center 
vortices which might percolate at the deconfinement transition.\cite{Hugo}  
The connection to the linear Coulomb or covariant gauges might be provided by
the Gribov ambiguity which suggests the infrared dominance of the Coulomb
field or the ghost correlations respectively.\cite{Gri78}
The deconfinement transition seems
harder to understand, but the different realizations of the center symmetry
should play an obvious role in these gauges also.

Recently, a complete truncation scheme for the 2--point Green's functions of
Landau gauge QCD, {\it i.e.}, the gluon, ghost and quark propagators, was 
established employing the present knowledge on the structure of the 3--point
vertex functions. This is possible by systematically neglecting all 
explicit contributions from 4--point vertices to the propagator DSEs as well
as in the constructions of the 3--point vertices. The truncated set of DSEs for
the gluon and ghost propagators of the pure gauge theory was solved
simultaneously in a one--dimensional approximation.\cite{Sme97,Sme98} The 
analytically extracted infrared behavior of the solutions contradicts
earlier solutions to the gluon DSE which relied on neglecting ghost 
contributions completely.\cite{Man79,Hau96}  While the 
gluon propagator of the couped system vanishes for small spacelike
momenta, the apparent contradiction with the earlier DSE
studies can be understood from the observation that the previously neglected
ghost propagator now assumes an infrared enhancement similar to what was then
obtained for the gluon. These results, in particular, the infrared
dominance of ghost correlations in Landau gauge, were later confirmed
qualitatively, by studies of a further truncated set of DSEs, neither to depend
on the particular constructions of the 3--point vertices nor to rely on the
one--dimensional approximation.\cite{Atk97,Atk98}  

A simultaneous solution to the propagator DSEs that includes the quark
propagator in the same scheme is presently under way.\cite{Ahl99}
In the quenched approximation, the solution to the gluon--ghost system gives
rise to an effective interaction in the gap equation for the quark--self
energy which is suppressed in the infrared as compared to a massless gluon
exchange. Nevertheless, dynamical chiral symmetry breaking can be described
quite well. This result is qualitatively different already from the existing
models of chiral symmetry breaking which basically all use interactions
equally strong or stronger than the massless gluon--pole in the
infrared.\cite{Haw94} 

Preliminary results obtained for the coupled system of ghost, gluon {\sl and}
quark propagators indicate towards little influence of the dynamical
quark--loops on the gluon and ghost propagators in the infrared. In
particular, the infrared dominance of ghost correlations seems to remain
unaffected. 

\section{The set of truncated gluon and ghost DSEs}

Besides all elementary 2--point functions, {\it i.e.}, the quark, ghost and
gluon propagators, the DSE for the gluon propagator also involves the 3-- and
4--point vertex functions which obey their own DSEs. These equations involve
successively higher n--point functions and are neglected in the present
scheme.\cite{Sme98}

The ghost and the transverse gluon propagator,
$ D_{gh}(k) = -{G(k^2)}/{k^2} $ and  $\;  D_{gl}^T(k) = {Z(k^2)}/{k^2}  $
respectively, are parameterized by two invariant functions $G$ and $Z$.
In order to arrive at a closed set of equations for these functions, first,
a Slavnov--Taylor identity (STI) for the ghost--gluon vertex was derived
from the usual Becchi--Rouet--Stora invariance. This STI can be resolved to
express the vertex in terms of $G$ and $Z$ additionally employing the
hermiticity of ghosts in Landau gauge, and thus the symmetry of the vertex,
while neglecting irreducible 4--ghost correlations herein, in accordance with
the truncation of the propagator DSEs.\cite{Sme98} Such constructions are
generally not unique. Undetermined transverse terms in this vertex can be
neglected also, however, as they correspond to a non--trivial 
ghost--gluon scattering kernel.\cite{erratum} These transverse terms will
become important in going beyond the present truncation scheme. Interesting 
constraints on their form arise from next--to--leading order perturbative
results.\cite{Wat99} Without explicit 4--gluon vertices the present scheme is
incomplete at 2--loop level anyway, however.

With the particular ghost--gluon vertex at hand, the construction of the 
3--gluon vertex follows standard procedures.\cite{Bar80} Its full Bose
symmetry is manifest and additionally possible transverse terms are
subleading in the infrared as well as in the perturbative limit.\cite{Sme98} 

An angle approximation has to be employed in the numerical solution of
the resulting system of coupled integral equations representing the truncated
gluon--ghost DSEs. The infrared behavior can then be discussed analytically,
the leading asymptotic form being $G \sim (k^2)^{-\kappa}$ and  $Z \sim
(k^2)^{2 \kappa}$ with $\kappa \simeq 0.92 $ for $k^2 \to 0$.
This leading behavior of the gluon and ghost renormalization
functions and thus of their propagators is \emph{entirely} due to ghost
contributions. Qualitatively similar results were obtained using a bare
ghost--gluon vertex and neglecting the 3--gluon loop
completely.\cite{Atk97,Atk98} Compared to the Mandelstam approximation
without ghosts, in which the 3--gluon loop alone determines the infrared
behavior of the gluon propagator and the running coupling in Landau
gauge,\cite{Man79,Hau96} this shows the importance of ghosts. In contrast to
the infrared singular coupling obtained from the Mandelstam
approximation,\cite{Hau96} the result of the coupled gluon--ghost system
implies an infrared stable fixed point in the running coupling of the
non--perturbative subtraction scheme, defined by 
\[ \alpha_S(s) = \frac{g^2}{4\pi} Z(s) G^2(s)
     \to 
    \frac{16\pi}{9} \left(\frac{1}{\kappa} - \frac{1}{2}\right)^{-1}
    \approx 9.5 \; \; ,  \quad \mbox{for} \; \; s \to 0 \; . 
\]
The Euclidean gluon propagator of the present scheme violates
Osterwalder--Schrader reflection positivity.\cite{Sme98} It thus precludes
transverse gluon states to exist in the positive asymptotic Hilbert
space of physical states. This is interpreted as a manifestation of gluon
confinement.

\section{Comparison to lattice results}

The solutions to the coupled gluon--ghost DSEs compare quite well to 
recent lattice results.\cite{Hau98} These use various lattice implementations
of the Landau gauge condition to simulate the gluon\cite{Der98,Attilio}  
and the ghost propagator.\cite{Sum96} There are clear indications towards an
infrared vanishing gluon propagator to be seen on the lattice
also.\cite{Attilio} Especially the ghost propagator, however, is in compelling
agreement with the lattice data at low momenta. We therefore conclude that
present lattice calculations confirm the infrared dominance of the ghost
correlations in Landau gauge. The evidence for a violation of positivity for
transverse gluons, as observed on the lattice, has also tremendeously
increased recently.\cite{Ais97} 

In order to compare the running coupling of the present scheme
to lattice results as extracted from simulations of the
3--gluon\cite{All97,Bou98} and the quark--gluon vertex\cite{Sku98} one has to
account for the details in the definitions of the different schemes. 
The asymmetric schemes employed in most lattice calculations have thereby the
potential problem that the infrared divergent ghost correlations lead to
artificial infrared suppressions in the momentum dependence of the
asymmetrically defined couplings.\cite{Alk99}   

The ideal lattice calculation to compare to the present DSE coupling would
be obtained from a pure QCD calculation of the ghost--gluon vertex in Landau
gauge with a symmetric momentum subtraction scheme.\cite{Alk99,Attilio}

\section{Quark Propagator}

We have solved the quark DSE in quenched approximation.\cite{Ahl99}
The quark--gluon vertex therein obeys its non--Abelian STI with
the quark--gluon scattering kernel neglected. As a result, it explicitly
contains a ghost renormalization function,
$  \Gamma^\mu (p,q) = G(k^2) \Gamma^\mu _{\rm CP} (p,q)$
where $\Gamma^\mu _{\rm CP}$, is the Curtis--Pennington vertex.\cite{Cur90} 
In contrast to a naive Abelian approximation, this effectively leads to an
infrared vanishing coupling $\sim (k^2)^\kappa $ in the quark DSE. The
resulting interaction is peaked around a finite scale $\mu \approx 200$MeV.
In the infrared it is weaker than a massless gluon pole. This allows, {\it
e.g.}, to use the Landshoff--Nachtmann model for the pomeron in our
approach. From the solution we estimate a pomeron intercept of approximately
$ 2.7/$GeV
as compared to the typical phenomenological value of 2.0/GeV.\cite{Cud89}  

Besides this infrared suppressed interaction dynamical chiral symmetry breaking is
manifest in the quenched approximation. Using a current mass of $m(1{\rm
GeV})=6$MeV we obtain a constituent mass of approximately 170 MeV. In the
Pagels--Stokar approximation this yields approx. 50 MeV for the pion decay
constant. These numbers are quite encouraging, especially, since with the 
momentum scale fixed from the running coupling at the $Z$--mass  
no parameters other than the current quark mass were adjusted. 

Considering the quark loop in the gluon DSE one realizes that the quark loop
will produce an infrared divergence which is, however, subleading as compared to
the one generated by the ghost loop. In the latter there appear three ghost
renormalization functions in the numerator and one in the denominator leading
effectively to an infrared divergence of the order $(k^2)^{-2\kappa}$. In the
quark loop term there is only one factor $G$, and thus a divergence of type
$(k^2)^{-\kappa}$ is anticipated. It can be shown, however, that after
ultraviolet regularization and renormalization the coefficient of this 
infrared divergent
term vanishes. Preliminary results indicate that the influence
of the quark loop on the coupled system of gluons and ghosts 
is almost negligible, i.e. the propagators for gluons, ghosts
and quarks are nearly indistinguishable from their quenched counterparts.

\section*{Acknowledgments}

R.\ A.\ wants to thank David Blaschke, 
Frithjof Karsch and Craig Roberts for organizing this very valuable
workshop. He is grateful to Attilio Cucchieri and Sebastian Schmidt
for valuable discussions. We thank Peter Watson for e-mail
correspondence on perturbative results for the ghost-gluon vertex.

\noindent
This work was supported by DFG (Al 279/3-1) and the BMBF (06--ER--809).


\begin{thebibliography}{99}
\bibitem{erratum} 
 It was incorrectly asserted in Ref.\ \protect\cite{Sme98} that 
 undetermined transverse terms do not occur in this construction of the
 ghost--gluon vertex.
\end{thebibliography}
\end{document}